\documentclass[12pt,preprint]{aastex}

\slugcomment{Resubmitted Version, \today}

\shorttitle{Proper Motion of Pulsar B1800$-$21}
\shortauthors{Brisken et al.}

\begin{document}

\title{Proper Motion of Pulsar B1800$-$21}

\author{W. F. Brisken}
\affil{National Radio Astronomy Observatory,
   P.O. Box O, Socorro, NM 87801; wbrisken@nrao.edu}

\author{M. Carrillo-Barrag\'an\altaffilmark{1} and S. Kurtz}
\affil{Centro de Radioastronom\'\i a y Astrof\'\i sica, UNAM
    Morelia, Michoc\'an, M\'exico}

\and

\author{J. P. Finley}
\affil{Physics Department, Purdue University, 525 Northwestern Avenue
West Lafayette, IN 47907-2036}

\altaffiltext{1}{Instituto de Astronom\'\i a, UNAM, Apdo. Postal 70 - 264,
Ciudad Universitaria, M\'exico, D.F., CP 04510}

\begin{abstract}
  We report high angular resolution, multi-epoch radio observations of
  the young pulsar PSR B1800$-$21.  Using two pairs of data sets, each
  pair spanning approximately a 10 year period, we calculate the
  proper motion of the pulsar.  We obtain a proper motion of
  $\mu_{\alpha}=11.6 \pm 1.8$~mas\,yr$^{-1}$, 
  $\mu_{\delta}=14.8 \pm 2.3$~mas\,yr$^{-1}$,
  which clearly indicates a birth position
  at the extreme edge of the W30 supernova remnant.  Although this
  does not definitively rule out an association of W30 and PSR B1800$-$21,
  it does not support an association.

\end{abstract}


\section{Introduction}

Pulsar-supernova remnant associations are important for various
reasons.  When an association is confirmed, knowledge of the supernova
remnant (SNR) can
be used to constrain pulsar (PSR) parameters such as birth magnetic fields,
spin periods, luminosities, and beaming fractions.  Conversely, knowledge
of the pulsar can constrain SNR ages and distances, and illuminate remnant
evolution and uncommon morphologies (Kaspi 1996).
Although over 30 candidate pulsar-SNR associations are purported, fewer
than a dozen have been confirmed.   

Although confirmation of an association is desirable, the opposite ---
disproving an association --- can also be useful.  Pulsar proper
motions, which are often cited to support associations, can be a
definitive means to disprove candidate associations.  Even if an
association is disproven, the pulsar transverse velocity is useful
information. It can constrain stellar collapse models and contribute
to studies of pulsar velocity distributions and galactic electron
density distributions (e.g., Chatterjee et al.\ 2001).

\section{PSR 1800$-$21 and SNR W30}
PSR B1800$-$21 and SNR W30 have been the object of several studies,
including \citet{ode86}, \citet{kw90}, \citet{frail94}, and
\citet{finley94}.  An association between the two has been actively
discussed: Kassim \& Weiler suggest a weak association based on
similar age and position.  Frail et al.\ argue that the association is
unlikely, based on the absence of a pulsar wind nebula (owing to the
high transverse velocity if the supernova occurred at the center of
the remnant) and the discrepancy between OH and H\,I line profiles; and
Finley \& \"Ogelman present X-ray data and interpret the SNR
morphology as an interaction with the ambient medium, which might
salvage the association.

Pulsar B1800$-$21 was discovered by \citet{cl86} in their
high-radio-frequency survey for young and millisecond pulsars.  It is
bright (S$_{\mathrm{1400MHz}}$ = 7.6 mJy) ranking 70 of 1162 pulsars with known
S$_{\mathrm{1400MHz}}$ as per version 1.24 of the ATNF pulsar database
\citep{man05}.  It is also young, with a characteristic age
$\tau_{\mathrm{c}} \equiv P/2\dot{P} = 15.8$~kyr, ranking 31 of 1560
pulsars known to date (Manchester et al.\ 2005).  
It is well established that for
many young pulsars the characteristic age is a poor indicator of true
age.  A non-zero spin period at birth would over-estimate the pulsar age,
while a braking index smaller than the nominal value of 3 would
under-estimate the age.  Nevertheless, the characteristic age of B1800$-$21
is in rough agreement with the SNR age (15 -- 28~kyr) under the
assumption of a blast energy of 10$^{51}$ ergs (Finley \& \"Ogelman
1994).  Adopting 15.8~kyr for both objects, and assuming that the
supernova occurred at the geometric center of the remnant, the pulsar
would need an extreme transverse velocity of $\sim$1700~km~s$^{-1}$ in
the south-west direction to reach its present position.

The dispersion measure (DM) distance currently gives the best estimate of 
the pulsar's distance.  The Taylor \& Cordes (1993) Galactic electron
density model yields 3.9~kpc  for the pulsar's DM=233.99 pc~cm$^{-3}$.
The newer Cordes \& Lazio NE2001 model \citep{cl02} yields a similar value
of $3.84^{+0.39}_{-0.45}$~kpc.\footnote{All uncertainties reported reflect the 
most compact interval containing 68.3\% of the probability.}

W30 is a roughly spherical SNR, about 50$'$ in diameter, with PSR
B1800$-$21 lying to the south-west of the geometrical center (see
Fig. 1).   \citet{kw90} suggested a distance of $6\pm1$ kpc  to the SNR,
based on association with HII regions.  At this distance, the 50$'$ size
corresponds to 80~pc.  Revised models of the galactic rotation curve
\citep{bb93} indicate that the HII regions, and by inference the SNR,
are about 4.8~kpc distant. \citet{finley94}, assuming an initial blast
energy of $10^{51}$ ergs, find a distance range of 3.2 -- 4.3~kpc.

The 5.3~kpc distance originally reported for the pulsar \citep{cl86}
was sufficiently close to the (then) $6\pm1$ kpc SNR distance to encourage the
idea of an association.  Refinements to the pulsar and the SNR distances
have revised both distances downward, keeping hopes for an association alive.
In particular, the $3.39^{+0.39}_{-0.45}$~kpc distance estimate 
for the pulsar is in
reasonable agreement with the 3.2 -- 4.3~kpc distance estimate for the SNR.

With the goal of ascertaining the relative motion of the pulsar with
respect to the SNR,  we undertook a program to measure the pulsar proper
motion.  To date the only measurement of this proper motion is the
nondetection of $\mu_{\alpha} = 18 \pm 30$~mas~yr$^{-1}$,
$\mu_{\delta} = 400 \pm 470$~mas~yr$^{-1}$ by \citet{zou05} using
pulsar timing.

\section{Observations and Data Reduction}

New observations of the pulsar were obtained with the 
NRAO\footnote{The National Radio Astronomy Observatory is
  a facility of the National Science Foundation operated under
  cooperative agreement by Associated Universities, Inc.} 
  Very Large Array (VLA) on 2002
February 8 at 21~cm and on 2005 February 5 at 3.6~cm.  Each of these
observations of PSR B1800$-$21 was paired with a data set from the VLA 
archive observed in a similar mode.  The
2002 (21~cm) data were taken in spectral line mode, and were paired with line
data from program AV201 taken on 1993 January 18. The 2005 (3.6~cm) data were
taken in continuum mode and were paired with continuum data from
program AF244 taken on 1993 March 14.  A summary of the observational
parameters for each of the four data sets is provided in
Table~\ref{tab:observations}.

Each of the four data sets was calibrated and imaged individually (see
\S\S \ref{sec:contmode} and \ref{sec:linemode} below).  
The two pairs of data sets (one of line
data, the other of continuum) provide two independent estimates of the
pulsar's proper motion using independent techniques.

\subsection{Continuum Data} \label{sec:contmode}

The AF244 and AC774 3.6~cm observations of PSR B1800$-$21 were directly
referenced to a standard VLA calibrator, B1748$-$253 or J1751-2524,
allowing an absolute position determination.

Data reduction followed normal procedures, with several exceptions.
The AF244 data were taken in B1950 coordinates while the AC774 data
were taken in J2000.0 coordinates.  To render the two data sets fully
comparable we performed several additional operations on the
AF244 data.  First, when needed, the coordinates were transformed
from B1950 to J2000.0 using the AIPS task UVFIX.  Next, 
UVFIX was used a second time to correct for the UT1-UTC difference
and to correct for a time-stamping error common to all VLA
observations.  Finally, the coordinates of the phase calibrator
(B1748$-$253) were updated from their 1993 values to the improved
position used for the 2005 observations (J1751$-$2524).  We used the
VLA calibrator manual\footnote{Available at
  http://www.aoc.nrao.edu/$\sim$gtaylor/calib.html} J2000 position
17$^h$51$^m$51.263047$^s$ $-$25$^\circ$24$'$00.060610$''$ with an
expected accuracy of $<10$~mas.

AF244 observed the pulsar once during source rise and once during
transit.  As a third step, the source-rising scan was omitted in order
to match the {\it uv} plane coverage more closely to that of the
transit-only AC774 observation.  The final image noise and resolution
for both observing dates are given in Table~\ref{tab:observations}.

\subsection{Spectral Line Data} \label{sec:linemode}

The AV201 and AC629 programs both observed at 21~cm in the A-configuration
using 15 spectral channels spanning a 23.5~MHz band in two polarizations.
Due to an apparent front-end filter misconfiguration during the AV201
observations, only spectral channels 8 through 14 were used; the
remaining channels were highly attenuated.  For the AC629 reduction,
channels 2 through 14 were retained.  Channels 1 and 15 were omitted
because the bandpass was steeply sloped there, possibly introducing
non-closing errors.  Self-calibration was performed for both data sets.

Wide-field imaging was used to image the pulsar and eight additional
sources found within the primary beam.  Five of these sources were
point-like, and served to define a reference frame in which
(following McGary et al. 2001) a precise proper motion could be
determined.  All nine sources detected are listed in Table~\ref{tab:lband}.

\section{Proper Motion Analysis}

The two continuum data sets have a time baseline of 4346.08 days
(11.90 yr).  Given the positions listed in
Table~\ref{tab:params}, we calculate proper motions in right
ascension ($\alpha$) and declination ($\delta$) of $\mu_\alpha = 11.4
\pm 3.4$~mas~yr$^{-1}$ and $\mu_\delta = 13.7 \pm 3.5$~mas~yr$^{-1}$.  The
two spectral line data sets have a time baseline of 3317.84 days (9.08~yr).  
The source positions of these two observations imply
proper motions of $\mu_\alpha = 11.7 \pm 2.2$~mas~yr$^{-1}$ and
$\mu_\delta = 15.7 \pm 3.1$~mas~yr$^{-1}$.

For the continuum data, the uncertainty in the pulsar coordinates
was determined by the measurement of the proper motion of an additional
calibrator source, J1832$-$1035, which was observed during both epochs.
Although a null result was expected, a non-zero result was obtained.
This check source was 3.37 and 3.92 times more distant from the phase
calibrator J1751$-$2524 than was the pulsar, in right ascension and declination,
respectively.  The uncertainty in each coordinate of the pulsar proper
motion was derived from the observed proper motion of this check
source divided by this factor.  For the line data sets, the
uncertainty in the pulsar coordinates is a statistical error from the
coordinate system defined by the five point sources.

The agreement between the two independent proper motion estimates for
the pulsar is reassuring, and suggests that a single, combined proper
motion is warranted.
The signal-to-noise ratio weighted combined proper motion is 
$\mu_\alpha = 11.6 \pm 1.8$~mas~yr$^{-1}$
and $\mu_\delta = 14.8 \pm 2.3$~mas~yr$^{-1}$.  
The total proper motion is thus 
$\mu_{\mathrm{total}} = 18.7^{+1.4}_{-1.5}$~mas~yr$^{-1}$ 
in the direction $38.1^\circ$ east of north.

These 3.6~cm continuum observations offer the best measurement of the
absolute position of the pulsar on the sky.  The J2000.0 coordinates of
the pulsar are determined to be $18^h03^m51.4105^s$,
$-21^{\circ}37'07.351''$ at MJD 51544.0 (2000 January 1) with error of
about 10~mas in each coordinate.  Relevant parameters of the pulsar
are listed in Table~\ref{tab:params}.

\section{Discussion}

The newly measured proper motion, together with an adopted distance,
can be used to determine the transverse velocity of the pulsar.  The
NE2001 distance, $D = 3.8 \pm 0.4$~kpc, was used to compute a
transverse velocity of $347^{+57}_{-48}$~km~s$^{-1}$.  This
calculation includes a correction of 0.64~mas~yr$^{-1}$ for the
combined effects of the Sun's peculiar motion and the differential
Galactic rotation.  The net effect of this correction is that the LSR
transverse speed is 8~km~s$^{-1}$ greater than its apparent speed.
This speed is well within the normal range of young pulsar speeds
(see, e.g. Arzoumanian et al. 2002 and Brisken et al. 2003) which 
suggests that the DM distance estimate is reasonable.   
  
Independently of the distance estimation, an assumed pulsar age,
together with the measured apparent proper motion, allows the
calculation of the pulsar birthplace coordinates.  In particular,
assuming 15.8 kyr for B1800$-$21, and a proper motion of
$\mu_{\mathrm{total}} = 18.7^{+1.4}_{-1.5}$~mas~yr$^{-1}$ $38.1^\circ$
east of north, its birthplace in J2000.0 is $\alpha_0 =$ 18$^{\mathrm
  h}$ 03$^{\mathrm m}$ 38.0$^{\mathrm s}$ $\pm$ 2.3$^{\mathrm s}$,
$\delta_0 = -21^\circ 41' 18.2'' \pm 41.6''$.


The characteristic age for the pulsar is based purely on the observed
pulse period and its derivative and is a good proxy for the true age
given two conditions: (1) spindown is due to the magnetic dipole
radiation leading to braking index $n=3$, and (2) the initial spin
period was much less than the current spin period.  

Livingstone et al.
(2005) note that all braking index measurements made give values less
than the nominal $n=3$.  Only very young pulsars have such
measurements so it might be safe to assume that B1800$-$21 has $n <
3$.  In fact, an index as low as 2 is reasonable, which would double
its true age.  If so, the birth site would be even further from the
W30 remnant, and would nearly coincide with the HII region G8.14+0.23
(IRAS 17599$-$2148) and the dark cloud, traced by mid-IR emission
(see Fig.~1).  We note that this H\,II region is probably unrelated to
the pulsar-SNR complex; radio recombination line velocities suggest 
that it is kinematically distinct from the other HII regions found
near the remnant \citep{lockman89}.  

Faucher-Giguere \& Kaspi (2006) have tabulated initial spin periods
that have been estimated for nine young pulsars.  Four have values less
than or about 30~ms, four more are between 50 and 90~ms, and one
exceptional case (PSR~J0538+2817) is quite long, at 140~ms.  Based
solely on these values it seems unlikely that the true age is less
than half the characteristic age.  However, serious selection effects
have likely biased this sample.  Through population modeling
Faucher-Giguere \& Kaspi (2006) conclude that a wider range of birth
spin periods is actually likely.  Assuming a normally-distributed birth
spin period,  they report a distribution with $\mu=300$~ms
and $\sigma=150$~ms.  Given that the current spin period of B1800$-$21
is 134~ms, this distribution is clearly not useful in estimating $P_0$
for this pulsar.  Thus the characteristic age is likely correct to a
factor of 2; claiming an age range much smaller than this is not
well-justified.

It is clear that the pulsar did {\it not} originate at the geometric
center of W30, as seen in the \citet{kw90} image.  First, the
birthplace coordinates are far-removed from the SNR center for any
assumed age.  Second, the pulsar is currently at a position
approximately 106$^\circ$ west of north (in equatorial coordinates)
with respect to the center of W30.  The proper motion, at 38$^\circ$
east of north, is more nearly {\it toward} the center of W30, rather
than away from it. 

A more sensitive image of the W30 remnant has been reported by
\citet{bgg06} (see Fig.~1).  This image shows W30 to be the bright,
eastern part of a larger, non-thermal nebula.  Despite the more
extended radio continuum emission seen in the Brogan et al. image,
the pulsar birth position lies outside of the radio remnant (see also
the three color image of Brogan et al.).

\citet{finley94} suggested that confinement by inhomogeneous, dense,
star-forming gas may have shaped a complicated, asymmetric remnant,
with the pulsar lying near one edge.  The ambient interstellar medium,
as traced by thermal 8~$\mu$m radiation, clearly suggests an extensive
interaction with the remnant.  Moreover, as conjectured by Finley \&
\"Ogelman, the interstellar gas at lower Galactic longitudes probably
does absorb X-ray emission from the remnant, resulting in the
morphology that they reported.  However, the {\it MSX} image shows
relatively little emission at Galactic latitudes north of the pulsar
position, implying less (or at least cooler) cloud material in this
direction. Hence, one would naively expect the SNR to expand toward
northern latitudes in addition to or instead of toward higher Galactic
longitude, as seen in the 90~cm image.  Although we cannot
definitively rule out some pathological SNR morphology, the inferred
birth position of the pulsar does not lend support to an association.

The Brogan et al. (2006) 90~cm image also shows the newly-discovered SNR
G8.31$-$0.09 (seen in Fig.~1 as the cyan and yellow contours
approximately 0.2$^\circ$ below the pulsar positions).  The proper
motion we measure rules out this remnant as the B1800$-$21 birth site.
In addition, the range of possible birth locations do not coincide
with any interesting portion of the TeV source HESS J1804$-$216
\citep{ahar06}.

Associations between young pulsars and supernova remnants are
attractive as it is widely accepted that a single supernova can
produce such a pair. Some SNR / PSR associations are quite secure,
such as PSR J0538+2817 and SNR S147 \citep{klh03}.
However, for both PSRs B1800$-$21 and B1757$-$24 the
geometric coincidence and lack of contradictory data initially led to
speculation of association with SNRs W30 and W28, respectively. In the
case of B1757$-$24, a proper motion upper limit suggests insufficient
velocity to allow for an association \citep{tbg02}.
In general, the lack of an
associated radio pulsar with a known SNR is not surprising for several
reasons: not all supernovae result in pulsars; not all pulsars are
oriented for favorable detection at Earth, and some pulsars are
intrinsically too faint to be seen. Young pulsars without shell-type
SNRs are more mysterious.  The Crab nebula, created in year 1054 with
PSR B0531+21, is a pulsar-powered nebula without a well-defined SNR,
possibly indicating a very low energy supernova
\citep{fkcg95}.  The time since the supernova, the density structure of the
ambient ISM, and the mass of the progenitor star greatly affect the
visibility of SNRs.  The lack of associations with either PSR
B1800$-$21 or SNR W30 should not be alarming.

\acknowledgments

M. Carrillo-Barrag\'an thanks the Academia Mexicana de Ciencias for support
through its Summer of Research Program, during which the first
part of this work was done.  We thank D. Frail and C. Brogan for useful 
discussions, and C. Brogan for providing the radio-IR image of Figure 1.

\clearpage

\begin{figure}
\epsscale{.80}
\plotone{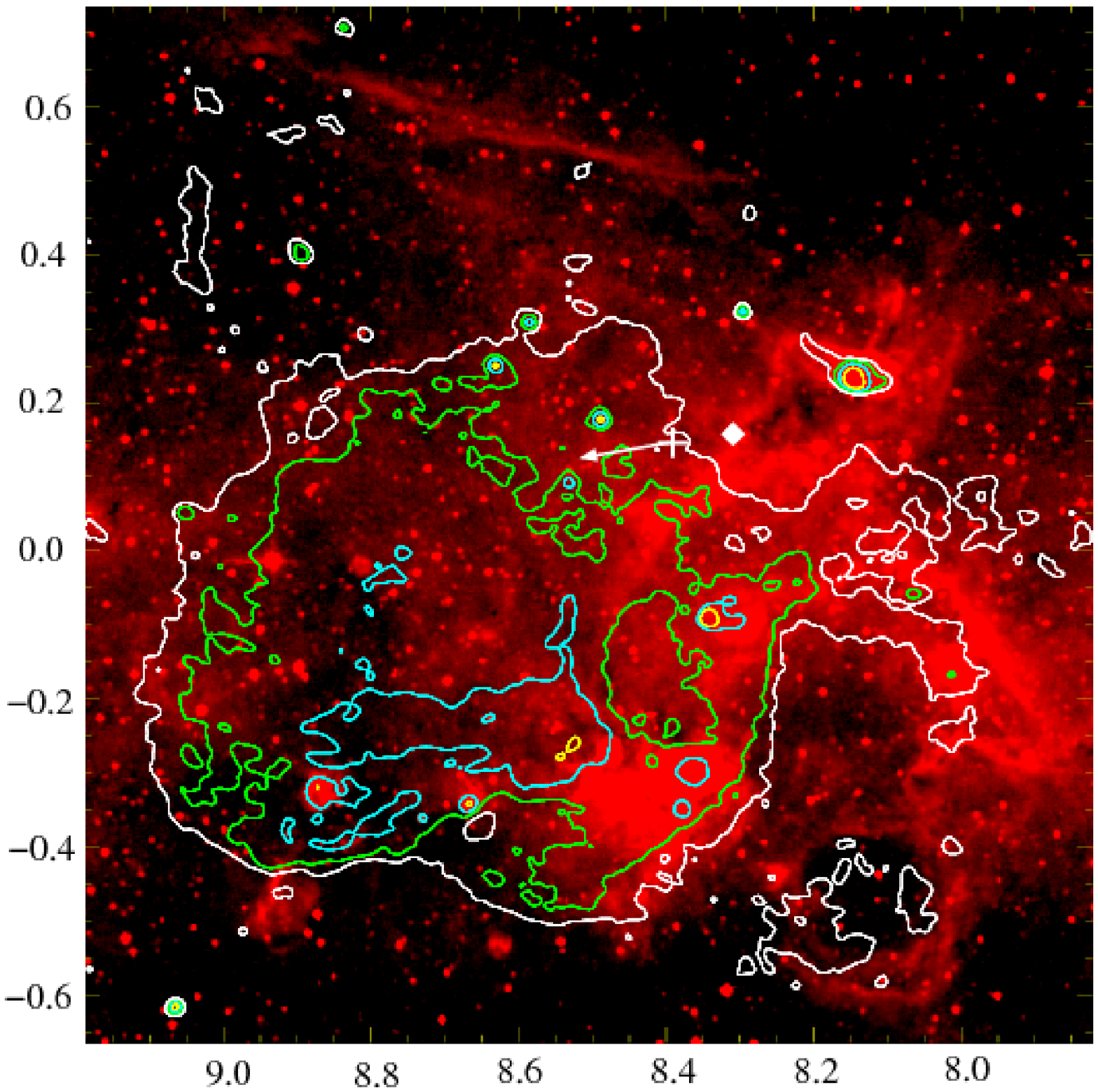}
\caption{Axes are Galactic latitude (vertical) and longitude (horizontal), 
in decimal degrees.
The red image (in logarithmic scale) shows infrared emission in 
the 8~$\mu$m band, as recorded by {\it MSX}.  The contours show the 90~cm
emission of the SNR; levels (in mJy~beam$^{-1}$) are 25 (white), 50 (green),
100 (cyan), and 200 (yellow). The white contour corresponds approximately 
to a 5$\sigma$ level.
The cross shows the J2000.0 pulsar position; the diamond shows the inferred 
birth position (see text, \S 5).  The arrow originating at the pulsar's position
points in the direction of its proper motion.
Both correlations and anti-correlations are seen between the mid-infrared
and the radio continuum emission, indicating extensive interaction between 
the SNR and the ambient interstellar medium.
\label{fig1}}
\end{figure}

\clearpage

\begin{deluxetable}{lcccccl}
\tablewidth{0pt}
\tablecaption{Observational summary \label{tab:observations}}
\tablehead{
    \colhead{Date}
  & \colhead{Program}
  & \colhead{VLA}
  & \colhead{$\lambda$}
  & \colhead{Synthesized}
  & \colhead{Image Noise}
 \\
  & \colhead{Code}
  & \colhead{Config.}
  & \colhead{(cm)}
  & \colhead{Beam}
  & \colhead{($\mu$Jy~beam$^{-1}$)}
  }

\startdata
1993 March 14   & AF244 & B   & 3.6$^{\dagger}$ & $1.29'' \times 0.68'' \; + 2.3^\circ$ & 48 \\
2005 February 5 & AC774 & BnA & 3.6$^{\dagger}$ & $0.68'' \times 0.42'' \; -89.0^\circ$ & 41 \\
1993 January 18 & AV201 & A   & 21$^{\ddag}$    & $2.05'' \times 1.08'' \; +4.5^\circ$  & 91 \\
2002 February 8 & AC629 & A   & 21$^{\ddag}$    & $3.03'' \times 1.22'' \; -28.3^\circ$ & 276 \\
\enddata
\tablenotetext{\dagger}{Observations made in continuum mode.}
\tablenotetext{\ddag}{Observations made in line mode.}
\end{deluxetable}

\clearpage

\begin{deluxetable}{lccccl}
\tablewidth{0pt}
\tablecaption{Sources in the L-band field of view \label{tab:lband}}
\tablehead{
    \colhead{Source}
  & \colhead{R.A.}
  & \colhead{Dec.}
  & \colhead{$\theta_{\mathrm{sep}}$}
  & \colhead{$S_{1460\mathrm{MHz}}^{\dagger}$}
  & \colhead{Comment}
 \\
  & \colhead{(J2000)}
  & \colhead{(J2000)}
  & \colhead{(arcmin)}
  & \colhead{(mJy)}
  &
}
\startdata
B1800$-$21    & $18^h03^m51^s.41$ & $-21^{\circ}37'07''.2$ &  0   &  7.0 & \\
1           & $18^h04^m38^s.03$ & $-21^{\circ}47'07''.6$ & 14.8 & 20.5 & \\
2$^{\ddag}$ & $18^h04^m20^s.42$ & $-21^{\circ}31'58''.8$ &  8.5 & 12.4 & 3 extended components \\
3$^{\ddag}$ & $18^h03^m55^s.56$ & $-21^{\circ}31'42''.4$ &  5.5 & 12.7 & jet structure \\
4           & $18^h03^m57^s.71$ & $-21^{\circ}21'51''.7$ & 15.3 & 10.9 & point-like galaxy core \\
5           & $18^h03^m38^s.65$ & $-21^{\circ}22'37''.0$ & 14.8 & 63.1 & \\
6$^{\ddag}$ & $18^h03^m01^s.39$ & $-21^{\circ}48'11''.0$ & 16.0 &  ?   & 30$''$ size structure \\
7           & $18^h02^m58^s.91$ & $-21^{\circ}37'30''.5$ & 12.2 & 13.8 & \\
8           & $18^h04^m14^s.97$ & $-21^{\circ}45'42''.1$ & 10.2 &  2.5 & \\
\enddata
\tablenotetext{\dagger}{Flux densities are corrected for primary beam attenuation.}
\tablenotetext{\ddag}{Three sources were not used as astrometry reference sources but
were imaged and used in the self-calibration process.}
\end{deluxetable}
%

\clearpage

\begin{deluxetable}{lr}
\tablewidth{0pt}
\tablecaption{B1800$-$21 parameters \label{tab:params}}
\tablehead{
    \colhead{Parameter}
  & \colhead{Value}
}
\startdata
Right ascension, $\alpha$ (J2000)                        & $18^h03^m51.4105(10)^s$      \\
Declination, $\delta$ (J2000)                            & $-21^{\circ}37'07.351(10)''$ \\
Reference epoch (MJD)                                    & 51544 \\
Proper motion in $\alpha$, $\mu_{\alpha}$ (mas yr$^{-1}$) & 11.6(1.8) \\
Proper motion in $\delta$, $\mu_{\delta}$ (mas yr$^{-1}$) & 14.8(2.3) \\
Total proper motion, $\mu_{\mathrm{total}}$ (mas yr$^{-1}$) & $18.7^{+1.4}_{-1.5}$ \\
Dispersion measure, DM (pc cm$^{-3}$)                    & 233.99 \\
NE2001 DM Distance, $D$ (kpc)                            & $3.84^{+0.39}_{-0.45}$ \\
Transverse velocity, $v_{\perp}$ (km s$^{-1}$)           & $347^{+57}_{-48}$ \\
Pulse period, $P$ (s)                                    & 0.13362 \\
Pulse period derivative, $\dot{P}$ ($10^{-13}$)          & 1.3410 \\
Characteristic age, $\tau_{\mathrm{c}}$ (kyr)            & 15.8 \\
Birth right ascension$^{\dagger}$, $\alpha_0$ (J2000)    & $18^h03^m38(2)^s$ \\
Birth declination$^{\dagger}$, $\delta_0$ (J2000)        & $-21^{\circ}41'18(42)''$ \\
\enddata
\tablenotetext{\dagger}{Birth location assumes $\tau_{\mathrm{c}}$ accurately reflects
the pulsar age.}
\end{deluxetable}

\clearpage

\end{document}